\pdfoutput=1

\documentclass[
 aip,
 apl,
 reprint,
]{revtex4-1}

\usepackage{graphicx}
\usepackage{upgreek}

\begin{document}

\preprint{AIP/123-QED}

\title[Sample title]{Versatile scanning tunneling microscopy with 120\,ps time resolution}

\author{Christian \surname{Saunus}}
\email{saunus@physik.rwth-aachen.de}
\author{Jan Raphael \surname{Bindel}}
\author{Marco \surname{Pratzer}}
\author{Markus \surname{Morgenstern}}
\affiliation{II. Institute of Physics B and JARA-FIT, RWTH Aachen University, D-52074 Aachen, Germany}

\date{\today}

\begin{abstract}
We describe a fully ultra-high vacuum compatible scanning tunneling microscope (STM) optimized for radio-frequency signals. It includes in-situ exchangeable tips adapted to high frequency cabling and a standard sample holder, which offers access to the whole range of samples typically investigated by STM. We demonstrate a time resolution of 120\,ps by using the nonlinear $I(V)$-characteristic of the surface of highly oriented pyrolithic graphite. We provide atomically resolved images in pulse mode related to a spatially varying nonlinearity of the local density of states of the sample, thus, demonstrating the possible spatial resolution of the instrument in pulse mode.
Analysis of the noise reveals that changes in the tunneling junction of 50\,pA are dynamically detectable at 120\,ps time resolution.
\end{abstract}


\maketitle


The STM is one of the few tools that enables imaging with atomic resolution.
However, it is complex to combine its unmatched spatial resolution with
a high time resolution, since the measurement bandwidth of common high-gain current
amplifiers does typically not exceed a few kilohertz at a sufficient noise level.\cite{Houselt2010}
In order to overcome these limits, Nunes and Freeman combined an STM with the pump-probe technique
known from time resolved optical probing.
They examined the cross correlation current of two photo gated voltage pulses applied to the sample \cite{Nunes1993}
and demonstrated 20\,ps time resolution combined with 1\,nm spatial resolution.\cite{Khusnatdinov2000}
Their high temporal resolution relies on the careful guiding of the voltage pulses
on the sample towards the tunneling junction by the use of impedance matched transmission lines. Thus, their approach is not versatile with
respect to sample preparation.
Recently, S. Loth {\textit et al.} detected electron-spin relaxation with atomic resolution using an all-electrical pump-probe scheme and
a standard, thus, versatile STM system.\cite{Loth2010} In their work the shortest measured relaxation times were 50\,ns using a probe pulse width of 100\,ns.
Later, another group reported an all electrically achieved time resolution of 300\,ps with atomic resolution, but again
using ex-situ prepared strip-lines on the sample side.\cite{Moult2011}

\begin{figure}
\includegraphics[width=1\linewidth]{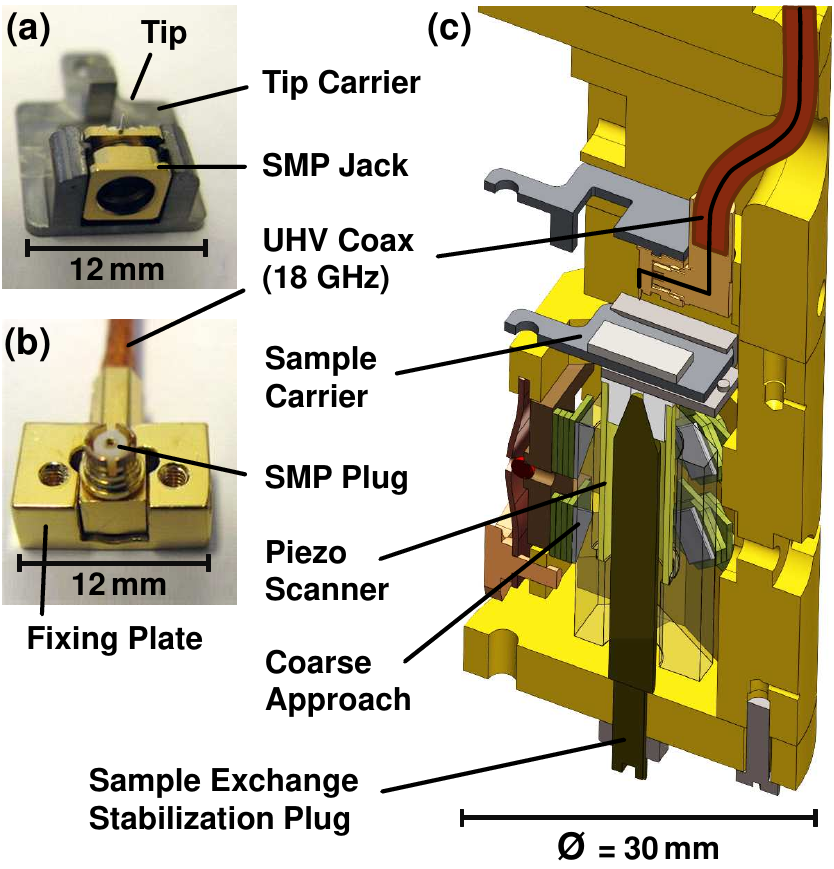}
\caption{\label{stm}
(a) SMP jack fixed to a modified standard tip carrier. Short tip is glued directly to the jack as marked. (b) SMP plug terminating the RF coaxial cable as used for applying the voltage pulses. Plug is glued to a fixing plate which is screwed to the STM body. (c) Schematic cross-section through the STM with main components marked.
}
\end{figure}

Here, we present a home-built STM head using radio-frequency (RF) optimized tips,
which achieves a time resolution of 120\,ps while maintaining atomic resolution and relying on versatile sample holders.
The time resolved response detected by a lock-in amplifier is reproduced by a parameter-free simulation based
on the measured $I(V)$-curve of the tunneling junction.
While the presented experiments are performed at ambient conditions providing a resolution in current change of 50\,pA,
the STM is fully compatible with ultra-high vacuum (UHV), low temperature (5\,K)
and high magnetic fields since only slightly modified with respect to established designs.\cite{Mashoff2009}
Tunneling tips and samples are exchangeable in-situ and the sample holder is identical to that used for
conventional STM giving full access to the plethora of known UHV sample preparation methods as well as coating of the tip by ferromagnetic or anti-ferromagnetic
materials for the purpose of spin-polarized STM.\cite{Bode}

Figure \ref{stm}(c) shows a cross-section of the microscope scan head. The design is similar to a conventional Pan-design STM,\cite{Mashoff2009,Pan1993} but
features two conceptional differences.
Firstly, the position of tip and sample are exchanged. The sample holder is mounted on the piezo tube scanner on the bottom side and the tip holder is fixed rigidly on the upper side, such that we can use rigid coaxial wiring for the tunneling voltage applied to the tip which enhances the transmission of RF-frequencies.
To prevent the piezo tube from damage during sample exchange, there is a stabilization plug in the center of the tube as labeled in Fig. \ref{stm}(c).
When the coarse approach is fully retracted, the plug slides into a macor counter cone, which is glued directly to the sample holder and counteracts
the lateral forces during the exchange.

Secondly, the tunneling voltage is guided towards the tip by an RF-coaxial cable with cut-off frequency $f_{\text{-3\,db}} = 18$\,GHz.
Applying the voltage to the tip rather than the sample avoids complex RF-optimized sample designs as long as the sample is well grounded capacitively.
To allow in-situ tip exchange, the coaxial cable is terminated by the SMP-type right angle plug shown in Fig. \ref{stm}(b), which features a cut-off frequency of $f_{\text{-3\,db}} = 40$\,GHz. The plug is fixed to the back of the microscope body. The right angle counter jack, shown in Fig. \ref{stm}(a), is mounted on a
modified standard tip carrier with the short tunneling tip (2\,mm) glued to the center conductor of the jack.
The whole tip assembly can be exchanged as plug-in board in-situ using a wobble stick.
Noise measurements of an identical sample stage performed in UHV revealed a $z$-noise of 2\,pm RMS.

\begin{figure}
\includegraphics[width=1\linewidth]{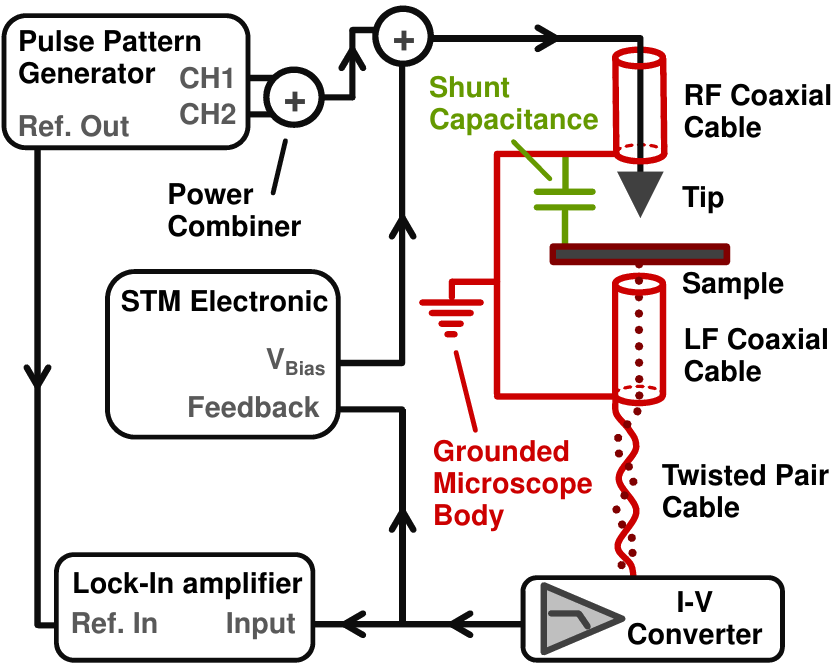}
\caption{\label{setup}
Setup for pulsed STM measurements: Pulse pattern generator (PPG) generates ``pump'' (CH1) and delayed ``probe'' (CH2) voltage pulses, which are summed by a power combiner (+). The time delay of CH2 is modulated with a reference frequency to enable lock-in detection. The second power combiner adds the bias voltage provided by the STM electronics. The RF coaxial cable guides the signal to the tip of the microscope which is unshielded on the last 2\,mm. The sample is connected via low frequency wiring to an $I\text{-}V$-converter with bandwidth 1\,kHz and gain $10^9$\,V/A. For frequencies exceeding the bandwidth of the preamplifier the circuit is shunted by the parasitic sample-ground capacitance (labeled ``shunt capacitance''). The signal of the $I\text{-}V$-converter is fed back into the STM electronics and simultaneously into the input channel of the lock-in amplifier. For pulse superposition measurements (Fig. \ref{100ps}) the lock-in output is displayed as a function of pump-probe delay.}
\end{figure}

The pulse measurement setup is shown in Fig. \ref{setup} consisting of a pulse pattern generator (PPG)\cite{ppg} and a lock-in amplifier \cite{sr830}. The PPG has two independent output channels which provide voltage pulses of up to 2\,V at 50\,$\Omega$ and a full width at half maximum (FWHM) measured to go down to 120\,ps (nominally 100\,ps). The channels can be internally delayed and are summed by a resistive power combiner at the output of the PPG. A second power combiner adds the bias voltage of the STM electronics to the voltage pulses before the signal is guided to the tunneling tip by the RF cable. The reflection of the pulses from the tunneling gap is mostly absorbed by the power combiner.\cite{suppl} The resulting tunneling current is first amplified by $10^9$\,V/A at a bandwidth of 1\,kHz and then analyzed by the feedback electronics and the lock-in amplifier simultaneously.
In pulse STM operation, the PPG sends two continuous pulse trains, a ``pump'' pulse train and a ``probe'' pulse train, to the junction, modulating the time delay between them with a fixed reference frequency to allow lock-in detection. During the first half of a modulation period the delay between pump and probe pulses is set to the time delay of interest, which will be swept during a time resolved measurement. During the second half period the time delay is set to a value at which the probe pulse is separated as far as possible from the pump.
The lock-in amplifier, thus, displays a value proportional to the average excess current generated by the probe pulses at the time delay of interest.
The phase of the lock-in is optimized to the resistive signal by setting it 90$^\circ$ off the phase
providing the maximum capacitive signal, which is recorded with a $5\pm 2$\,nm withdrawn tip.

\begin{figure}
\includegraphics[width=1\linewidth]{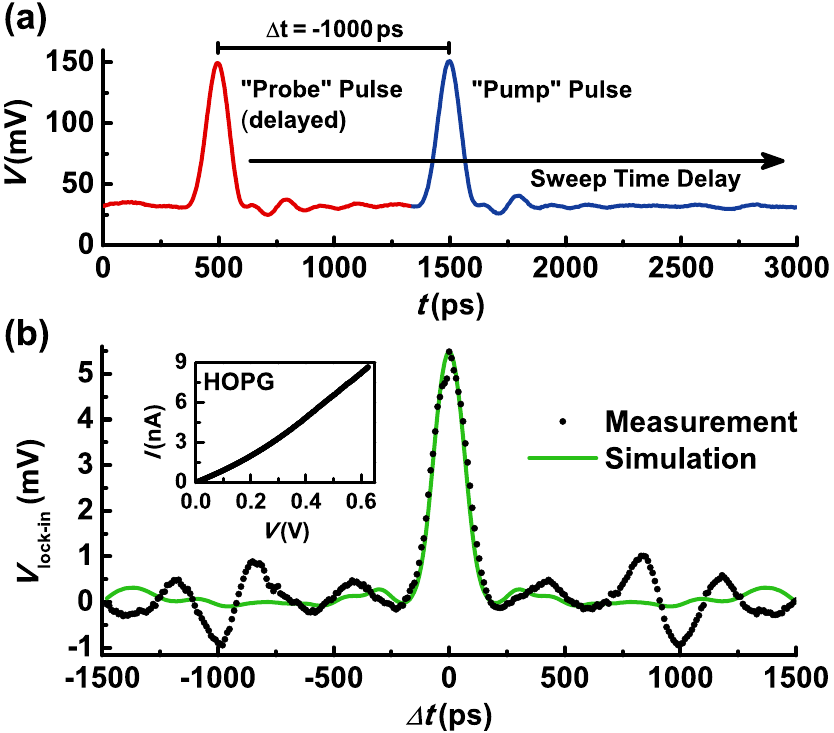}
\caption{\label{100ps}
(a) Voltage pulse train generated by pulse pattern generator (nominal amplitude 2\,V, nominal width 100\,ps) and combined with DC bias as measured by an oscilloscope, connected at 50\,$\Omega$ input impedance to the output of the second power combiner. The combined attenuation of both power combiners (21.6\,dB) results in 121\,mV pulse amplitude. The pulse train is repeated at 166\,MHz, while the ``probe'' pulse delay is modulated between the delay of interest $\Delta t$ and a time delay of 3\,ns at 1403\,Hz. $\Delta t$ is swept during measurements. The figure shows $\Delta t=-1$\,ns, i.e. the probe precedes the pump pulse. (b) Lock-in output $V_{\text{lock-in}}$ as a function of $\Delta t$ for an HOPG sample and a PtIr tip. Each measurement point is recorded after a waiting time of three lock-in time constants $\tau_{\text{lock-in}}=300$\,ms. Three measurement runs are averaged. $V_{\text{bias}}=54$\,mV, $I_{\text{set}}=500$\,pA. Solid line: Simulation of the lock-in output using the measured $I(V)$-characteristic of the tunneling junction (inset) and the pulse train shown in (a).
}
\end{figure}

The time resolution of a pump-probe measurement is limited by the width of the voltage pulses directly at the tunneling junction.\cite{Nunes1993} It cannot be measured directly by an oscilloscope without affecting the voltage drop across the junction. Thus, one has to use a different approach. Figure \ref{100ps}(b) shows the result of a pulse superposition measurement on highly oriented pyrolytic graphite (HOPG) under ambient conditions. The acquired lock-in voltage $V_{\text{lock-in}}$ is shown as a function of the time delay $\Delta t$ between the two pulse trains. The pulses as measured by an oscilloscope directly connected to the second power combiner (see Fig. \ref{setup}) are displayed in Fig. \ref{100ps}(a). They have a full width at half maximum (FWHM) of 120\,ps and an amplitude of 121\,mV (nominally 2\,V) at the 50\,$\Omega$ input impedance of the oscilloscope. The attenuation originates from the two resistive power combiners. The voltage amplitude at the tunneling junction is presumably larger due to its higher impedance. We estimate the factor by measuring the same dc-voltage with the oscilloscope and a 10\,M$\Omega$ input impedance voltmeter leading to a factor of 1.83.

The tunneling junction consists of a PtIr tip above a flat HOPG terrace produced by cleavage. The tip is stabilized at a bias voltage of $V_{\text{bias}}=54$\,mV and a tunneling current of $I_{\text{set}}=500$\,pA in order to characterize the junction by the $I(V)$ curve shown in the inset of Fig \ref{100ps}(b). The current rises non-linearly with increasing voltage, i.e. $I(V)$ offers a positive curvature.
Then, the PPG output is enabled and the lock-in value is recorded during a time delay sweep. The feedback loop of the STM electronics is kept enabled.
We apply a 300\,Hz low pass filter to the feedback current so that the feedback responds to the slow thermal drift and piezo creep, but not to the 1403\,Hz delay modulation frequency.

The non-linear $I(V)$-characteristic of HOPG leads to a gain in average current per cycle when pump and probe pulses start to overlap ($\Delta t \rightarrow 0$\,s) as seen in Fig. \ref{100ps}(b).\cite{Nunes1993,Moult2011}
In particular, the current at twice the voltage originating from full pulse overlap ($\Delta t = 0$\,s) is, due to the positive $I(V)$ curvature, larger than twice the current at the peak voltage of a single pulse. Thus, the time averaged current of this ``overlap pulse'' is higher than the time averaged current of the separated pulses.
Since the average current of the separated pulses is subtracted by the lock-in amplifier, Fig. \ref{100ps}(b) only shows contributions from the excess current.

In order to estimate the time resolution, we calculate the expected lock-in output using the measured pulse shape of Fig. \ref{100ps}(a) and the measured correction factor of 1.83.
We take one period of the voltage train $V(t)$, convert it into the resulting current train $I(t)$ using the measured $I(V)$ curve (inset Fig. \ref{100ps}(b)) and determine the time average of this current train. This calculation is done twice for each time delay $\Delta t$, once with the desired time delay and once with the time delay set to maximum pulse separation. The mean value of these two average currents is chosen to be identical to $I_{\text{set}}$ and the difference between them describes twice the amplitude of the oscillating current signal at the input of the lock-in amplifier. The resulting $V_{\text{lock-in}}(\Delta t)$ curve is shown as solid line in Fig. \ref{100ps}(b). It exhibits excellent agreement with the measured data in the region of overlap (average deviation 4\,\% at $\Delta t=\pm 60$\,ps), but does not reproduce the additional oscillations which probably originate from remaining reflections of the voltage pulses within the circuit. Hence, we firstly deduce that the voltage pulses at the tunneling junction are nearly identical to the ones provided by the PPG, respectively, we establish a time resolution of 120\,ps. Neither the cabling inside the STM, the remaining inductance of the short tip, nor the poorly RF-matched sample does limit the time resolution further. Secondly, we deduce that the DC conductivity of the tunneling junction determines the time-resolved signal, which connects it to the combined density of states of tip and sample.\cite{Tersoff1985}\\
The noise level of the measurement depends on the current noise at the modulation frequency and the time constant of the lock-in detection chosen to be 300\,ms.
The former can be reduced from the measured 8.2\,pA at 2\,kHz bandwidth by about a factor of ten using low-temperature UHV STM.\cite{Mashoff2009}
At ambient conditions, we could reduce the voltage amplitude by a factor of five with respect to Fig. \ref{100ps}, i.e. 44.3\,mV pulse amplitude at the junction,
still getting a signal to noise (S/N) ratio of 3 for the lock-in value at pulse overlap.\cite{suppl} This implies that changes tunneling current of 50\,pA can definitely be resolved at 120\,ps time resolution,\cite{suppl} which most likely gets further improved at low-temperature.

\begin{figure}
\includegraphics[width=1\linewidth]{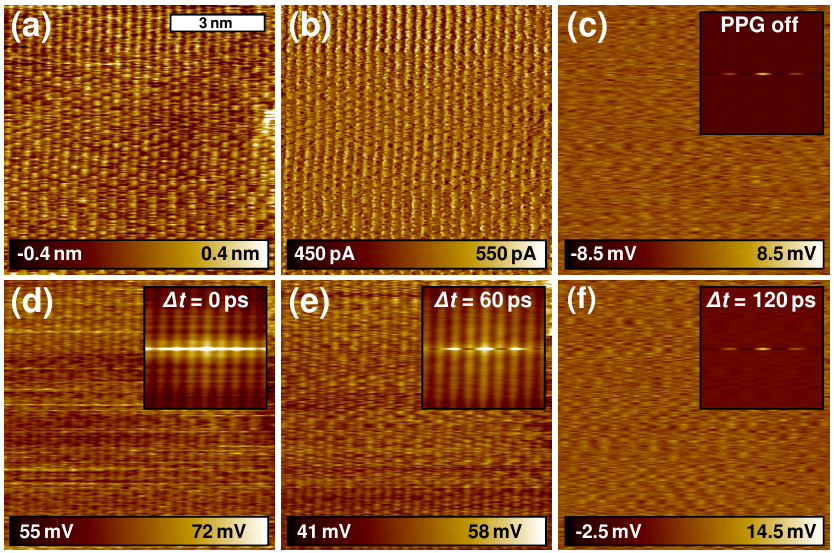}
\caption{\label{resolution}
(a) 8.6\,nm x 8.6\,nm constant-current image of HOPG with atomic resolution. $V_{\text{bias}}=48$\,mV, $I_{\text{set}}=500$\,pA. (b) Corresponding current image. (c) Simultaneously acquired lock-in data at 1403\,Hz reference frequency with a 3\,ms time constant exhibiting a 61\,mV RMS noise. (d)-(f) Lock-in data acquired the same way as in (c) but with enabled PPG using time delays $\Delta t$ as marked. Deduced peak voltage at the junction: 307\,mV. Measured pulse width: 133\,ps FWHM (150\,ps nominally). Repetition rate: 1.67\,GHz. The insets in (c)-(f) show corresponding autocorrelation images scaled to an identical contrast.
}
\end{figure}

Finally, we demonstrate atomic resolution in pulsed mode. Therefore, we image an 8.6\,nm x 8.6\,nm area of HOPG (Fig. \ref{resolution}(a)) and simultaneously acquire the lock-in output in PPG mode as a measure of the local non-linearity of the $I(V)$-curve. Since the current image (Fig. \ref{resolution}(b)) still shows atomic contrast, we firstly recorded a lock-in image without pulse application, which does not show atomic contrast (Fig. \ref{resolution}(c)) . Figures \ref{resolution}(d)-(f) show lock-in images on the same position with varying $\Delta t$ using pulses of a measured width of 133\,ps and pulse height at the junction of 307\,mV. For a time delay of 0\,ps and 60\,ps the images show atomic contrast in the real space images with a visibility of 5\,\% and 8\,\%, respectively, while the contrast disappears at 120\,ps , where the pulses hardly overlap. This result becomes even more striking within the autocorrelation images (insets), which exhibit atomic rows at $\Delta t=0$ ps and $\Delta t=60$ ps, but not at $\Delta t=120$ ps. Thus, the applied voltage pulses probe at an atomic scale and there is no obvious reason, that this would change for really time-varying signals.
The relatively large contrast of 5-8\,\% combined with the low error bar of 4\,\% in the simulation of Fig. \ref{100ps}(b) implies that spatial variations in the local curvature of $I(V)$ cause the contrast in line with previous arguments.\cite{Moult2011} Since spatial variations in $I(V)$ are well-known to be caused by the local density of states of the sample, \cite{Tersoff1985} we can conclude that we probe sample properties at 120\,ps time resolution. This is further substantiated in the supplement.\cite{suppl}

In conclusion, a home built RF-STM is presented, where the tip is directly connected to an RF-cable with 18\,GHz bandwidth. We demonstrate time resolution of 120\,ps while maintaining atomic resolution in pulsed mode. This is the highest temporal resolution with demonstrated atomic resolution. We maintained time resolution down to 45\,mV pulse amplitude implying a sensitivity for current changes down to, at least, 50\,pA. While the measurements where performed at ambient conditions, the STM is compatible with UHV, low temperature, and magnetic field. Importantly, the spatiotemporal resolution is achieved without any additional restrictions to the sample with respect to standard STM, thus giving atomically resolved access, e.g., to charge carrier dynamics or mechanical motion of nano membranes\cite{mashoff2010}.


\begin{acknowledgments}
We acknowledge helpful discussions with S. Loth, H. Bluhm, M. Liebmann, M. Kl\"aui, B. Beschoten and C. Stampfer and financial support from the DFG via SFB 917-A3 and Mo 858/8-2.

Copyright 2012 American Institute of Physics. This
article may be downloaded for personal use only. Any
other use requires prior permission of the author and
the American Institute of Physics. The following article
appeared in Appl. Phys. Lett. 102, 051601 (2013) and may be found at http://link.aip.org/link/?APL/102/051601.

\end{acknowledgments}

\end{document}